\documentclass[lettersize,journal]{IEEEtran}
\usepackage{amsmath,amsfonts}
\usepackage{algorithmic}
\usepackage{subfigure}
\usepackage{algorithm}
\usepackage{array}
\usepackage{textcomp}
\usepackage{stfloats}
\usepackage{url}
\usepackage{verbatim}
\usepackage{multirow}
\usepackage{booktabs}
\usepackage{graphicx}
\usepackage[table,x11names]{xcolor}
\usepackage{cite}
\usepackage{caption}
\usepackage{subcaption}
\usepackage{svg}
\usepackage{pgfplots}
\usepackage{tikz}
\usepackage{circuitikz}
\usepackage{hyperref}
\usepackage{xspace}
\usepackage{scalerel}
\usepackage{tikz}
\usetikzlibrary{svg.path}

\definecolor{orcidlogocol}{HTML}{A6CE39}
\tikzset{
  orcidlogo/.pic={
    \fill[orcidlogocol] svg{M256,128c0,70.7-57.3,128-128,128C57.3,256,0,198.7,0,128C0,57.3,57.3,0,128,0C198.7,0,256,57.3,256,128z};
    \fill[white] svg{M86.3,186.2H70.9V79.1h15.4v48.4V186.2z}
                 svg{M108.9,79.1h41.6c39.6,0,57,28.3,57,53.6c0,27.5-21.5,53.6-56.8,53.6h-41.8V79.1z M124.3,172.4h24.5c34.9,0,42.9-26.5,42.9-39.7c0-21.5-13.7-39.7-43.7-39.7h-23.7V172.4z}
                 svg{M88.7,56.8c0,5.5-4.5,10.1-10.1,10.1c-5.6,0-10.1-4.6-10.1-10.1c0-5.6,4.5-10.1,10.1-10.1C84.2,46.7,88.7,51.3,88.7,56.8z};
  }
}

\newcommand\orcidicon[1]{\href{https://orcid.org/#1}{\mbox{\scalerel*{
\begin{tikzpicture}[yscale=-1,transform shape]
\pic{orcidlogo};
\end{tikzpicture}
}{|}}}}
\usepackage{hyperref}

\begin{document}
\pagenumbering{arabic}

\title{ADS-IMC: \underline{A}ccelerating \underline{D}ata \underline{S}orting with \underline{I}n-\underline{M}emory \underline{C}omputation}

\author{Narendra Singh Dhakad \orcidicon{0000-0003-2848-1785} and Santosh Kumar Vishvakarma \orcidicon{0000-0003-4223-0077}

\thanks{Narendra Singh Dhakad was with Nanoscale Devices, VLSI Circuit and System Design (NSDCS) Lab, Department of Electrical Engineering, Indian Institute of Technology Indore, MP 453552, India. He is now with Intel Technology India Pvt. Ltd., Bangalore, India. This work was done during his PhD at the Indian Institute of Technology Indore.}

\thanks{Santosh Kumar Vishvakarma is associated with Nanoscale Devices, VLSI Circuit and System Design (NSDCS) Lab, Department of Electrical Engineering, Indian Institute of Technology Indore, MP 453552, India.}}

\maketitle

\begin{abstract}
Sorting is a fundamental operation across numerous computational domains. Traditionally, this process involves transferring data from main memory to a processing unit for sorting, followed by writing the sorted data back to memory. This conventional approach incurs substantial latency and energy overheads due to the extensive data movement between memory and processing components. To mitigate these overheads, this paper introduces novel architectures for executing sorting operations directly within the memory fabric, eliminating the need for off-chip data transfer. To our knowledge, this work represents the first exploration of in-memory sorting using 6T SRAM. The proposed architecture is designed to operate on data represented in the standard weighted binary radix format commonly used in digital systems. The proposed architecture achieves a significant 3.4$\times$ reduction in latency compared to memristor-based IMC sorting.
\end{abstract}


\section{Introduction}
\IEEEPARstart {S}orting is a fundamental operation in computing, serving as the backbone for various applications in fields such as databases \cite{database1,database2}, scheduling\cite{scheduling}, and scientific computing\cite{scientific}. Efficient sorting algorithms are critical for optimizing search operations, data retrieval, and processing, making them essential in a wide range of industries.  Sorted indexes in databases expedite search and retrieval. Sorting is necessary for packet scheduling in networking, which guarantees efficient data transfer. Sorting is used in scientific computing to organize massive datasets, which is necessary for precise simulations and analysis. The evolution of computing has forced the development of both software and hardware-based sorting methods, each with its distinct advantages and applications. 

Traditionally, sorting has been performed in software, with algorithms like QuickSort, MergeSort, BubbleSort, and HeapSort \cite{Softwarealgo} being widely used. Software-based sorting algorithms are flexible, easy to implement, and can be optimized for different types of data and computing environments. However, as data volumes continue to grow, the limitations of software-based sorting become apparent, particularly in terms of speed and energy efficiency \cite{datadepend}.

On the other hand, hardware-based sorting offers several advantages, particularly in scenarios where high-speed sorting is critical. Hardware sorting leverages parallelism, allowing multiple sorting operations to occur simultaneously, significantly reducing the sorting process's time complexity. This is particularly beneficial in real-time systems, where delays in sorting can lead to significant performance bottlenecks \cite{Memristor}.

The key difference between hardware and software sorting lies in their approach to parallelism and efficiency. While software sorting is generally executed sequentially, hardware sorting can be designed to perform multiple operations concurrently. This parallelism is achieved through specialized hardware components such as Field Programmable Gate Arrays (FPGAs) and Application-Specific Integrated Circuits (ASICs)\cite{FPGA1, FPGA2}.

In this work, we explore a Batcher-based \cite{batcher} in-memory sorting architecture utilizing 6T SRAM with the help of the compare and swap (CAS) block, a fundamental building block for many sorting networks. A CAS block is responsible for comparing and swapping two input values if they are out of order, ensuring that the output is a partially sorted set. This operation is repeated in parallel across multiple stages to achieve a fully sorted output.

A typical CAS block consists of a comparator and two multiplexers. The comparator evaluates the magnitude of the two inputs, while the multiplexers route the inputs based on the comparison result, effectively swapping the values if necessary. In the context of in-memory sorting, the CAS block is implemented directly within the memory array, allowing sorting operations to be performed without transferring data between memory and processing units. For in-memory sorting, the design of the CAS block is critical for the overall performance of the sorting network. It needs to be optimized for speed, area, and power consumption, especially when implemented in memory \cite{batcherlatency, batcherlatency2}. 

\begin{figure}[t]
\centering
\includegraphics[scale=0.6]{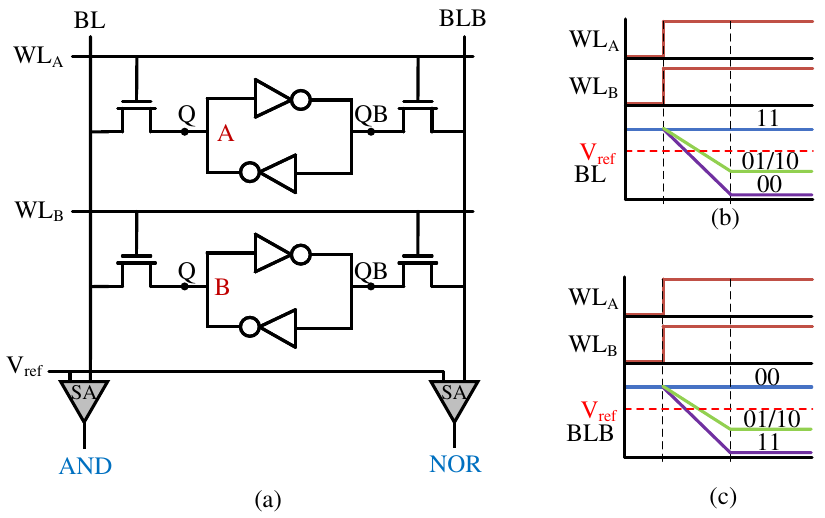}
\caption{(a) 6T SRAM-based IMC to implement logic over bitlines; (b) AND logic on BL; (c) NOR logic on BLB \cite{MainSRAM}.}
\label{arch8}
\end{figure}

In recent years, in-memory computing (IMC) has been explored for various data-intensive applications such as image classification, speech recognition, and matrix-vector multiplication \cite{MainSRAM, vishal}. IMC sorting is explored in \cite{Memristor} using RRAM. However, 6T SRAM is a prominent choice for hardware implementation due to its high speed, low latency, and established manufacturing processes \cite{6TSRAM}. In this paper, we explored the in-memory computing for sorting applications using SRAM. Here, we use 6T SRAM-based IMC to perform AND and NOR over bitlines (BL and BLB) as shown in Fig. \ref{arch8}. The rest of this article contains Section \ref{proposed}, which explains the proposed in-memory binary sorting designs. Section \ref{simulations} analyses the performance of the proposed design, simulations, and comparison of the proposed work with other works. Finally, a conclusion is drawn in Section \ref{conclusion}.



\begin{figure}[t]
		\centering
		\subfigure[4-bit comparator block]{\includegraphics[scale=0.2]{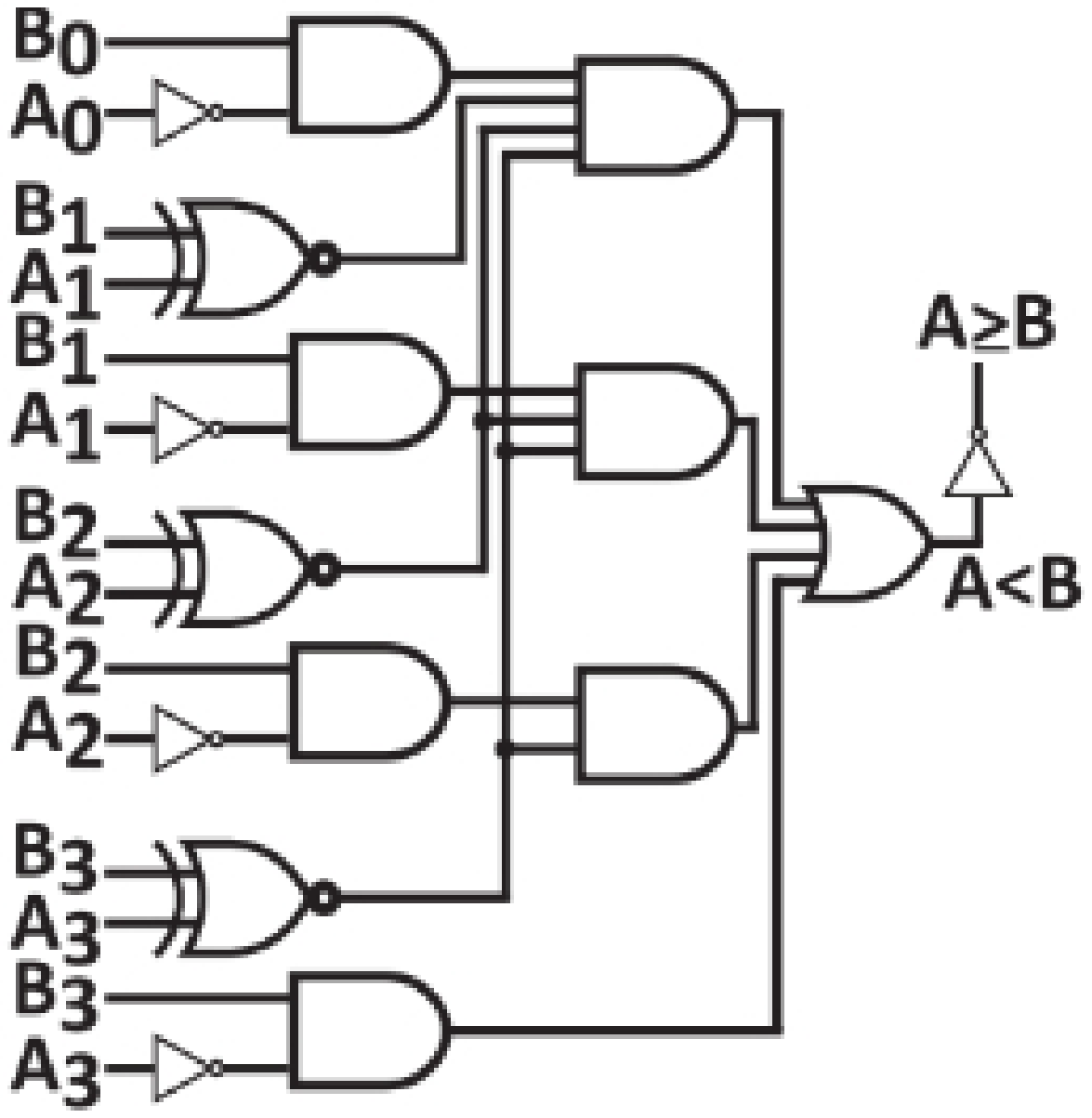}}
		\subfigure[multiplexer block]{\includegraphics[scale=0.2]{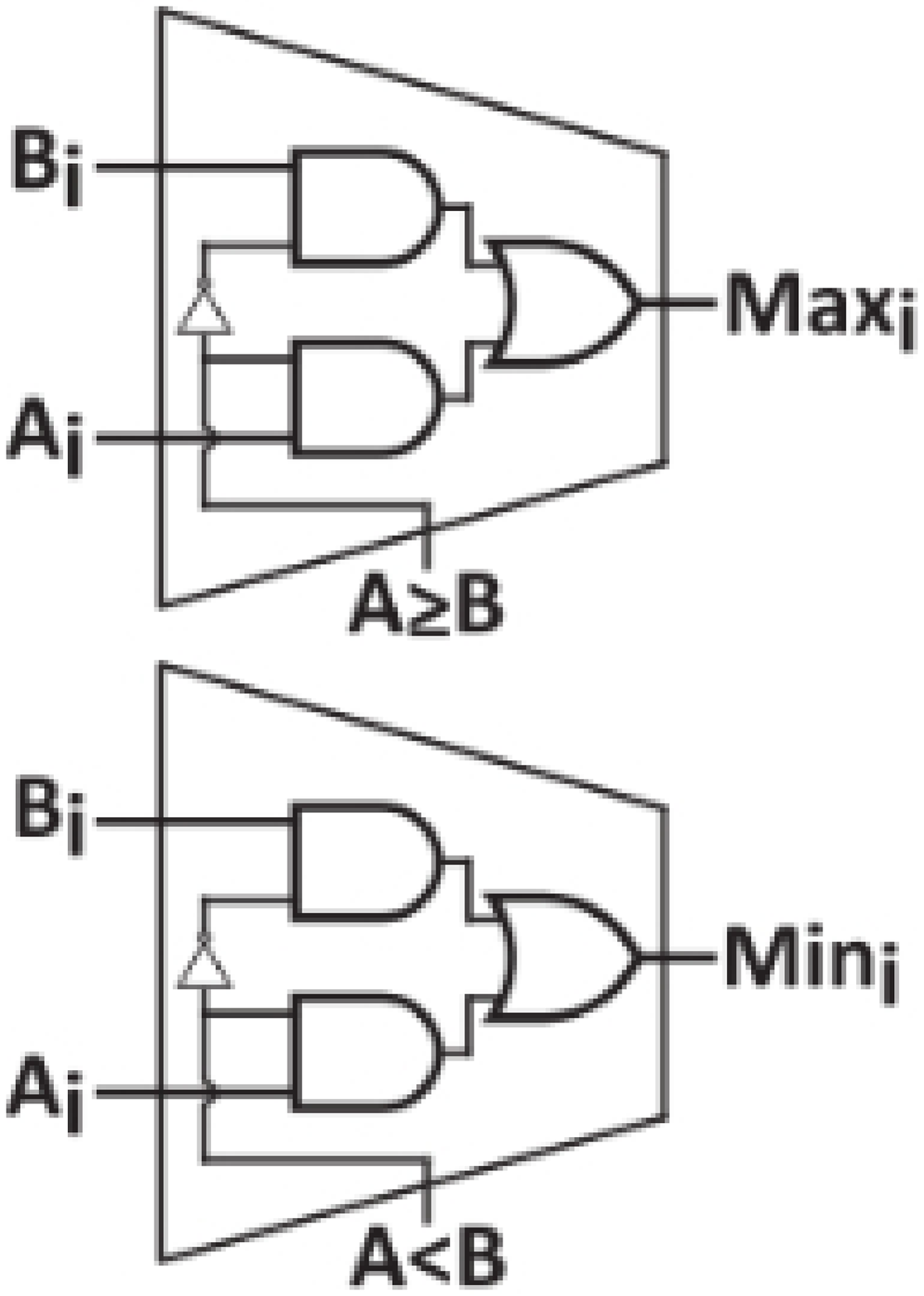}}
		\caption{Generic logic design \cite{Memristor}}
		\label{arch}
\end{figure}

\section{Proposed Sorting-in-Memory Architecture} \label{proposed}
\subsection{Compare and Swap Block}
The Compare and Swap block, which serves as the fundamental binary sorting unit, comprises one comparator and two multiplexers. Figure \ref{arch} shows the generic binary comparator and multiplexer blocks for 4-bit binary numbers. We use 6T SRAM IMC architecture to implement this comparator and multiplexers. The generic logic comparator typically relies on three and four input gates for computation, which is not feasible with the 6T SRAM-based IMC due to data flipping issue \cite{MainSRAM}. So, we used two input logic operations for the comparator and multiplexer implementation.

\begin{figure}[t]
\centering
\includegraphics[width=\linewidth]{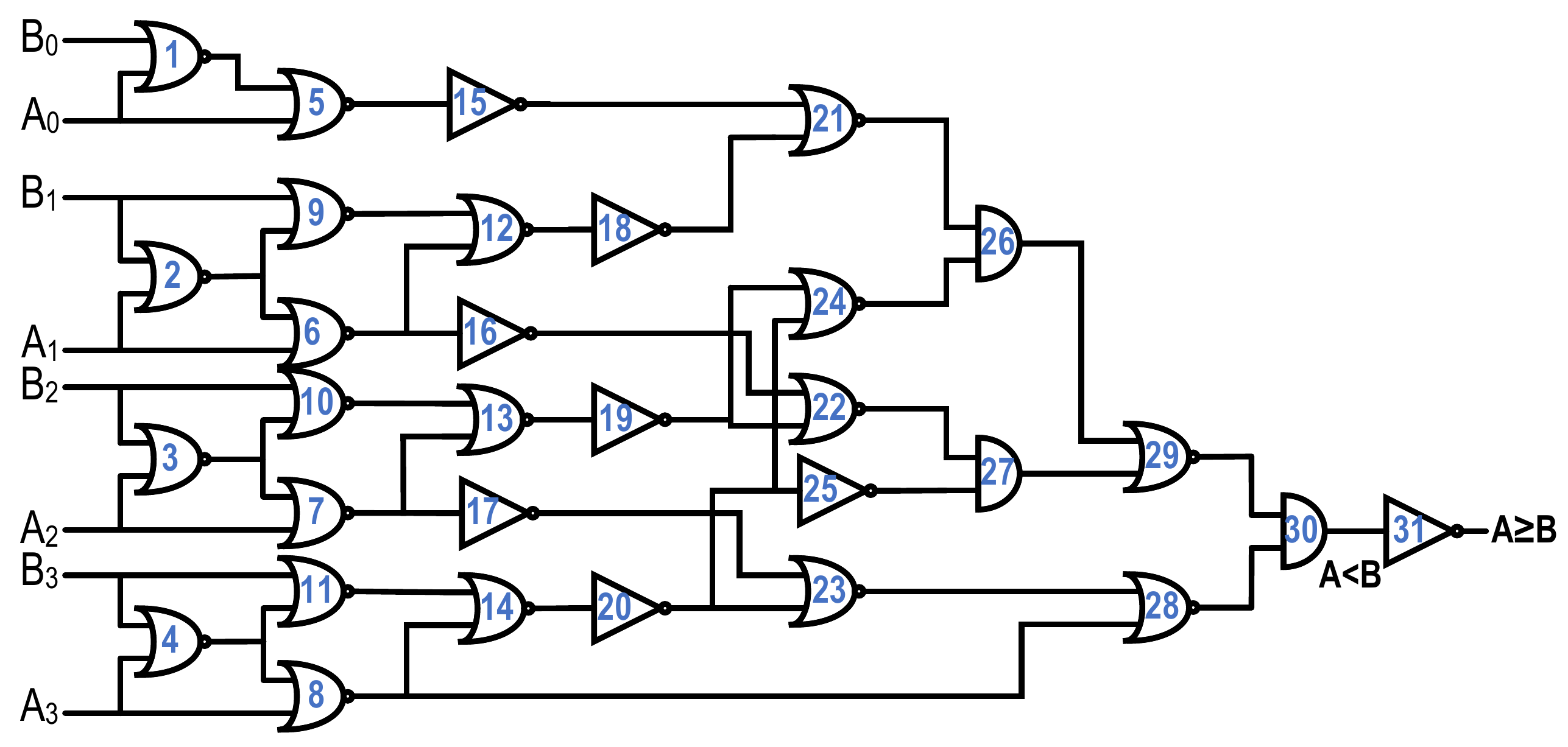}
\caption{Proposed logic design of a 4-bit comparator block}
\label{arch3}
\end{figure}

\begin{figure}[t]
\centering
\includegraphics[scale=0.25]{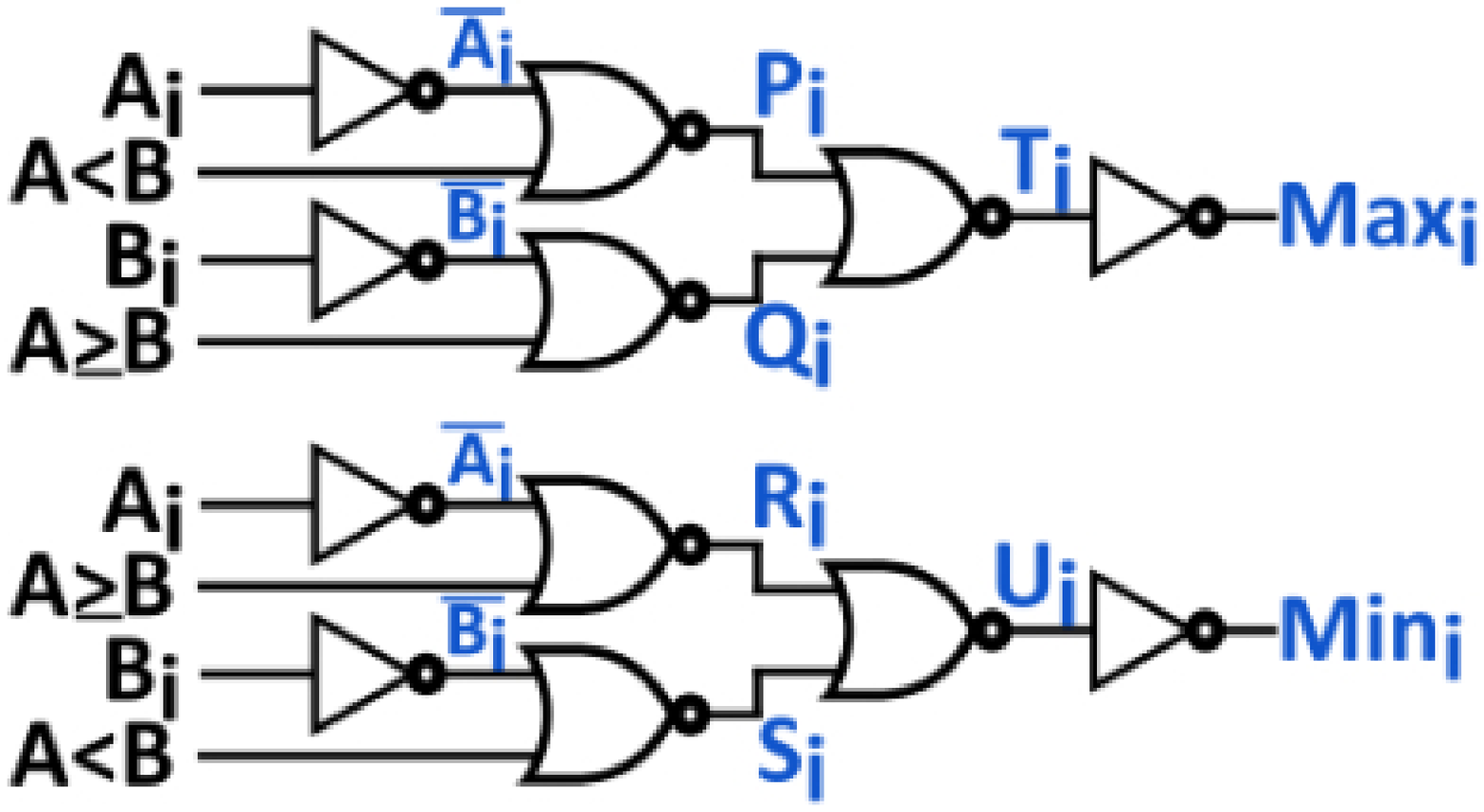}
\caption{Logic design of multiplexer block}
\label{arch4}
\end{figure}

\begin{figure}[t]
    \centering
    \subfigure[Comparator]{\includegraphics[scale=0.2]{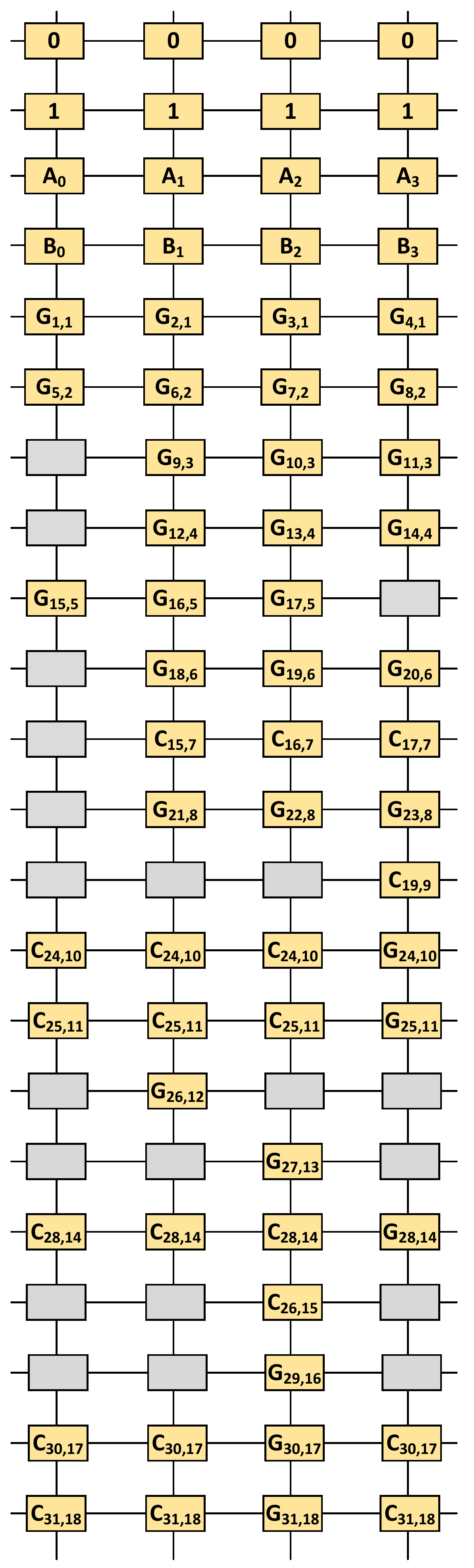}}
    \subfigure[Multiplexer]{\includegraphics[scale=0.2]{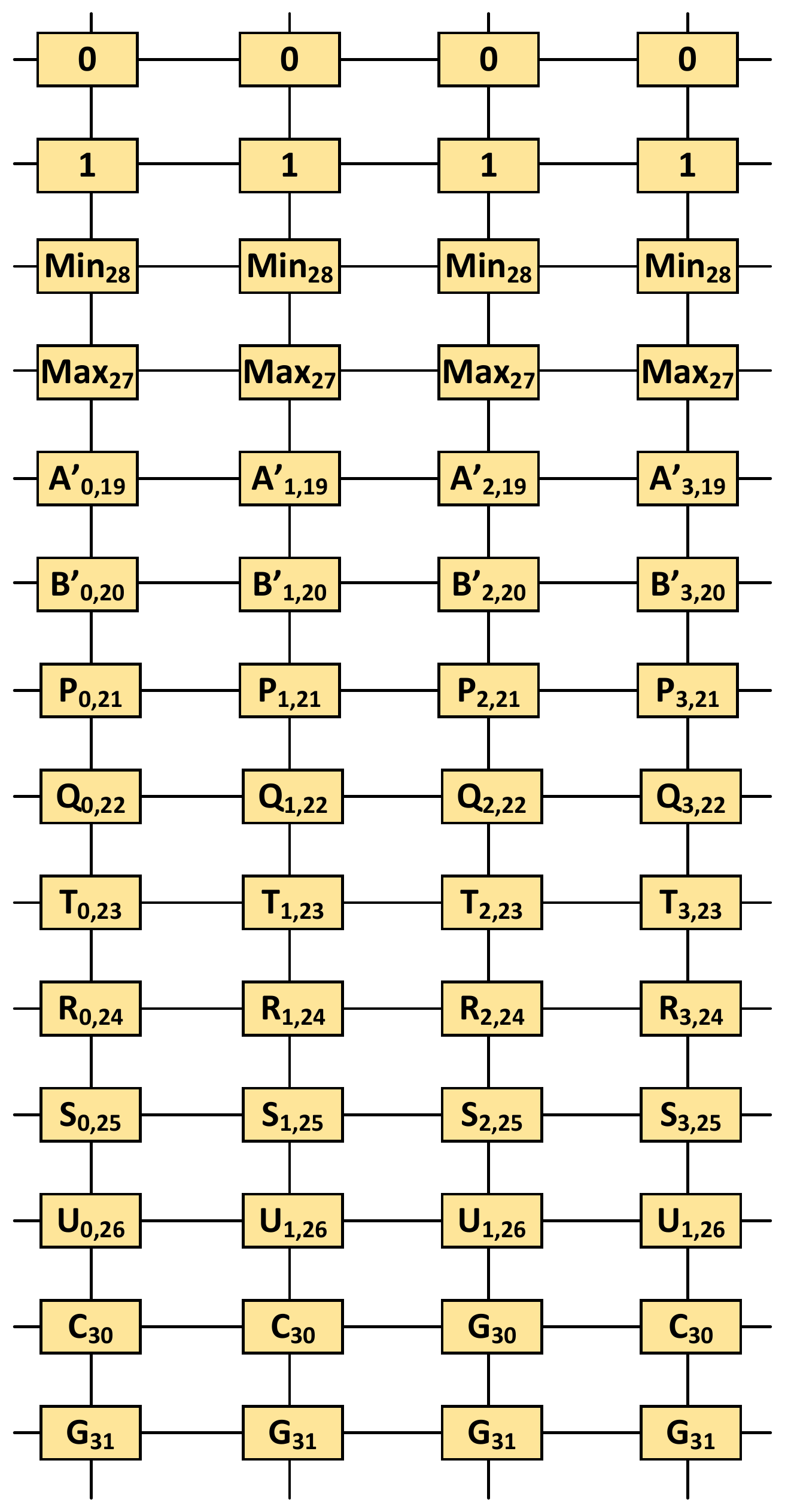}}
    \caption{Proposed in-memory computing architecture; where Gi is the output of the gates \& Ci represents a copy of Gi.}
    \label{arch5}
\end{figure}

Figures \ref{arch3} shows the logic for the comparator using two input gates, specifically NOR/AND-based logic, respectively, which is implemented using in-memory computing. we use an array of 4$\times$22 memory cells as shown in Figure \ref{arch5}(a). In this figure each memory cell is used once for comparator and once for multiplexer. The memory size can be reduced from 4$\times$22 to 4$\times$9 by reusing memory cells multiple times. The computation uses NOR, NOT, AND, and COPY operations. The first-row stores logic 0 and the second-row stores logic 1, which is necessary because the 6T SRAM IMC cannot perform NOT and COPY logic operations directly. By using logic 0 as one of the inputs during the execution of the NOR operation, the NOT logic of the other input is achieved. Similarly, to copy any value, an AND operation is performed using logic 1 as one of the inputs. For sorting two data having 4-bit each, we store data $A=A_{0}A_{1}A_{2}A_{3}$ in row 3 and data $B=B_{0}B_{1}B_{2}B_{3}$ is stored in row 4, initially. Every row after these initial four rows is the result of logical operations executed in a cycle, as shown in Figure \ref{arch5}. 

During write back, four types of data movements take place: a) Normal logical operations, either AND/NOR, with the data written back to the same column; b) Copying data to the adjacent right column; c) Writing the data obtained from the last column to all columns simultaneously in the same cycle; and d) Writing the data obtained from the third column to all columns simultaneously in the same cycle. During each cycle, only one of these operations is executed, and a 4$\times$1 multiplexer is used to select which operation will be carried out. Out of the previous four shifting operations, the last two operations (i.e. C \& D) are architecture-specific and need to store the operation results in all of the columns. 

In Figure \ref{arch5} bitcell is denoted as $G_{i,k}$, where i represents the output of the $i^{th}$ logic gate, and $k$ denotes the cycle number in which it is obtained. Correspondingly, $C_{i,k}$ indicates a copy of the $i^{th}$ gate in $k^{th}$ cycle. For instance, bitcell $G_{29,16}$ signifies that it stores the output of $29^{th}$ gate in the proposed two-input logic and is obtained during the $16^{th}$ cycle as shown in Fig. \ref{arch3}. 
The comparison result is achieved in the $3^{rd}$ column during the $17^{th}$ cycle at bit-cell $G_{30,17}$, and for subsequent operations, this value is copied to the rest of the columns of the same row within the same cycle. Additionally, the content of the $21^{st}$ row is inverted in the $22^{nd}$ row to get the $G_{31,18}$, which is utilized as the select inputs for the multiplexer block. The entire Compare block takes a total of 18 cycles.

The memory array used for the comparator is reused to implement the multiplexer logic (shown in Fig. \ref{arch4}), which significantly reduces the number of bit cells required and, consequently, the area utilization of the array as shown in Fig. \ref{arch5}(b). After obtaining the comparison results, the same 4$\times$22 array is reused for the multiplexing logic, with six rows of this array remaining untouched (Rows 1, 2, 3, 4, 21, and 22). Rows 3 and 4 store the initial data (A and B) to be sorted. The multiplexer takes a total of 8 cycles to perform the multiplexing, which is then swapped based on the multiplexer output during $27^{th}$ and $28^{th}$ cycle. The minimum value is stored in $3^{rd}$ row in the $28^{th}$ cycle, while the maximum value is stored in row 4 during the $27^{th}$ cycle. Here, IMC architecture takes a total of 28 cycles to sort the two values for 4-bit precision.

\subsection{Complete Binary Sorting Unit }
\begin{figure}[b]
\centering
\includegraphics[scale=0.25]{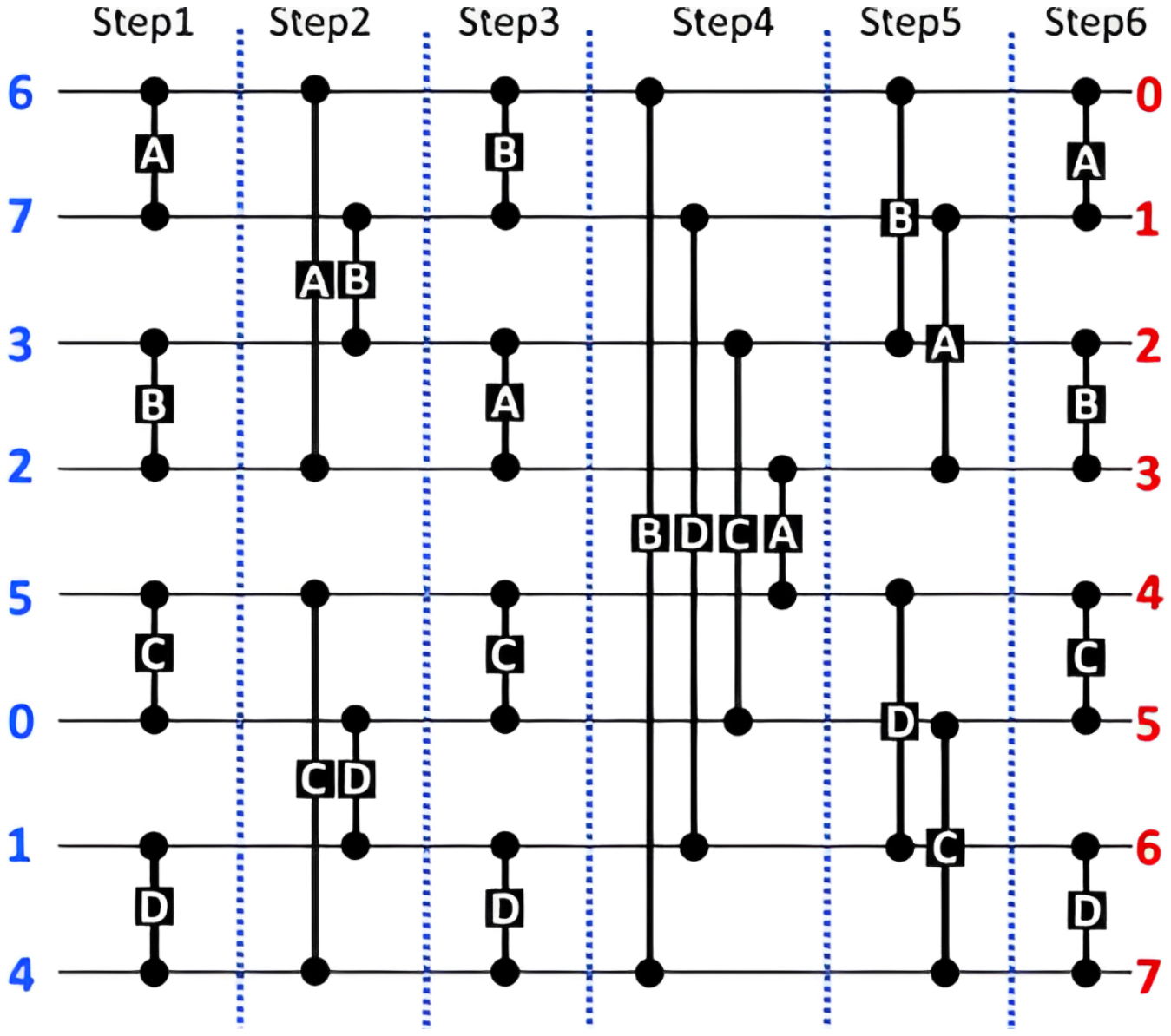}
\caption{8 input complete sorting unit.}
\label{arch7}
\end{figure}

In Bitonic sorting, the network repeatedly merges two sets of a given size to create a larger, sorted set of size N \cite{CAS}. Figure \ref{arch7} illustrates the CAS network used for an eight-input bitonic sorting process. As shown, the network consists of 24 CAS units. Generally, a bitonic sorting network with N inputs requires a specific number of CAS units. These CAS units can be organized into steps or stages, with each stage comprising N/2 CAS units that operate concurrently \cite{CAS}. Equation \ref{N_CAS} and \ref{N_steps} signifies the general expression for the number of CAS blocks required and several steps/stages required for a complete sorting unit.

\begin{equation}
N_{\text{CAS}} = N*(log_2N)*(1+log_2N)/4
\label{N_CAS}
\end{equation}
\vspace{-0.7cm}
\begin{equation}
N_{\text{Stages}} = log_2N*(1+log_2N)/2
\label{N_steps}
\end{equation}

Gupta et al. \cite{Memorypartition} introduced a memory partitioning technique to enhance in-memory parallelism. Following a similar approach, we divide the memory into several partitions, allowing different CAS operations to execute simultaneously within each stage of the bitonic CAS network. Figure \ref{arch7} demonstrates the implementation of an eight-input bitonic sorting network within memory, where the memory is segmented into four partitions labeled A, B, C, and D (indicated by black vertical lines in the bitonic network diagram). The number of partitions is determined by the number of CAS units that can operate in parallel.

Each partition holds two of the eight unsorted input data elements. The sorting process is divided into six stages, corresponding to the number of CAS groups. In the initial stage, the two inputs within each partition are sorted using the basic sorting operation described in Section \ref{proposed}(A). In the subsequent stage, the maximum value (i.e., the larger number) identified by the first stage’s sorting operation is transferred to another partition as required by the bitonic network configuration. With four memory partitions running in parallel across six stages, the total number of CAS operations required is 4$\times$6=24. The array size needed for the four partitions is equivalent to 16$\times$22 arrays to accommodate these operations. Two additional rows are used as temporary storage to transfer data across all four partitions and obtain new input values at each stage for comparison and swapping in the subsequent partition. At the end of one compare and swap operation, any two rows can be picked up for temporary storage. This process requires an extra 6 cycles for input provisioning, which adds up to 24 additional cycles in the complete sorting unit for 8 inputs system. For the N input sorting algorithm, Equation \ref{N_TemporaryRows} shows the number of temporary rows required for data swapping between the stages. Equation \ref{N_Cycles} shows the number of extra cycles required.

\begin{equation}
N_{\text{Temporary rows}} = N/4
\label{N_TemporaryRows}
\end{equation}
\vspace{-0.6cm}
\begin{equation}
N_{\text{Cycles}} = 3*N/4
\label{N_Cycles}
\end{equation}

\begin{table}[t]
\caption{\scshape Number of Operation Cycles}
\label{tab:example_table}
\centering
\begin{tabular}{l|c|c} \toprule
& \textbf{CAS Block} & \textbf{Single Stage CAS} \\ \midrule
\textbf{Operation} & \textbf{\# of Inputs = 2} & \textbf{\# of Inputs = 8}  \\ \midrule
\textbf{NOR}             & 14   & 84     \\
\textbf{NOT}             & 8    & 48     \\
\textbf{AND}             & 3    & 18     \\
\textbf{COPY}            & 3    & 42    \\ \midrule
\textbf{Total}           & \textbf{28}   & \textbf{192}  \\ \bottomrule    
\end{tabular}
\end{table}

Table \ref{tab:example_table} shows the number of operations cycles needed for every operation type for a CAS block, a complete sorting unit of input size 8, and an intermediate operation between two stages. 

\section {Performance Analysis} \label{simulations}

\begin{figure}[t]
\centering
\includegraphics[width=\linewidth]{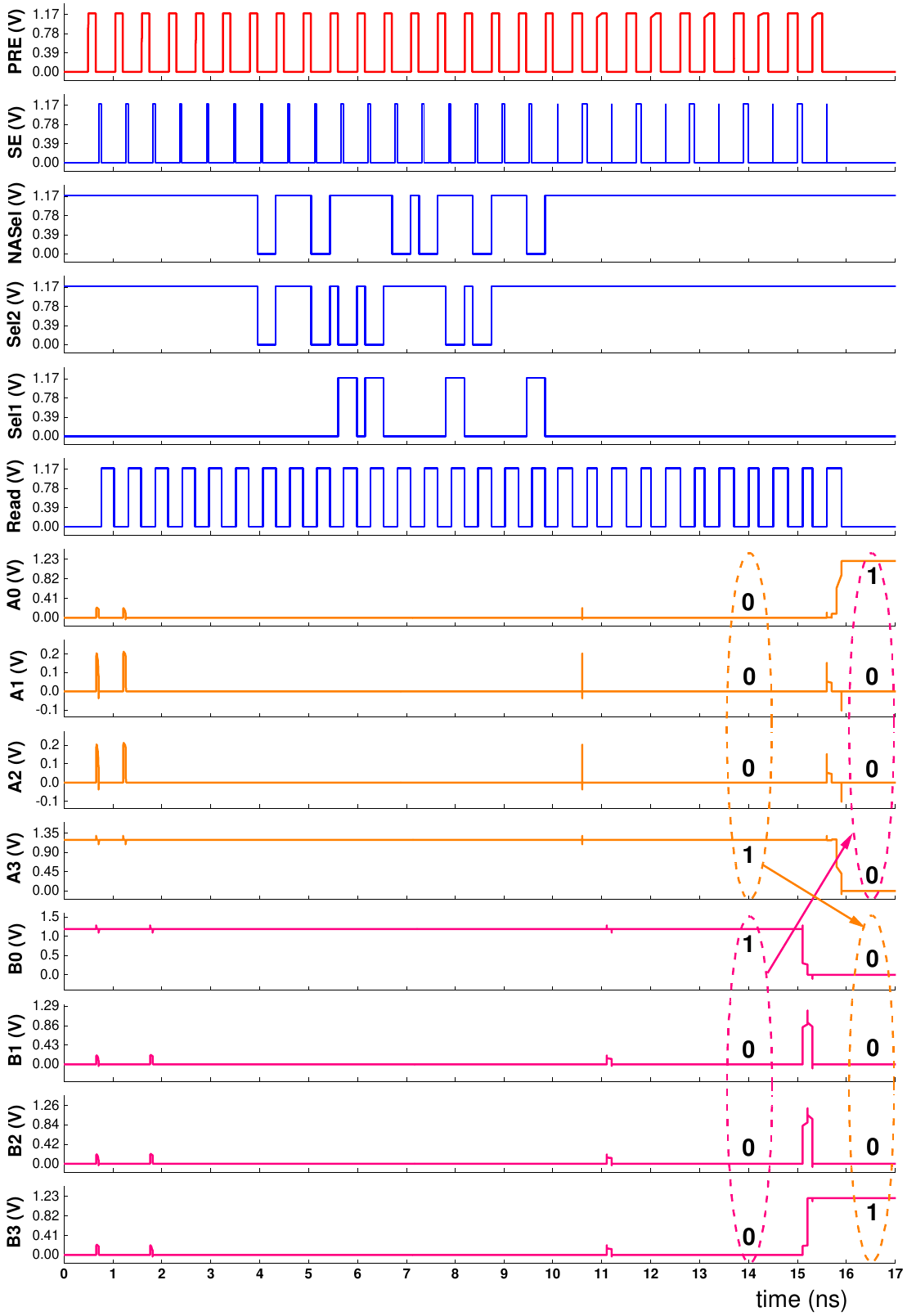}
\caption{Simulation waveforms of a CAS block, where $A=1000$ and $B=0001$}
\label{waveform}
\end{figure}

We implemented a 16$\times$22 crossbar of 6T SRAM using the Cadence Virtuoso at a 65nm technology node for performance evaluation. The latency of a single IMC operation is evaluated as 0.55ns. In a CAS block, for a network size of 2 and data width of 4 bits, a total of 28 cycles are needed. This makes CAS latency as 15.4ns (= 0.55$\times$28). As shown in Figure \ref{arch7} the CAS blocks A, B, C, and D execute in parallel. Data movement in between the stages takes 6 cycles extra, as one of the data in a particular CAS block is moved to another while other data is kept untouched. It takes only 4 numbers to interchange between the CAS blocks. Also, 2 temporary rows are required for this intermediate data movement. The latency of the complete binary sorting system for 8 inputs can make up a total of 192 cycles, which makes up to 105.6ns. The simulation waveform for a CAS block is shown in Fig. \ref{waveform}; here, input $A=1000$ (in binary radix) and $B=0001$ are considered for the waveform. Table \ref{tab:example_tabl} signifies the performance parameters of the proposed architecture.

\begin{table}[b]
\caption{\scshape Simulation Results}
\label{tab:example_tabl}
\centering
\resizebox{\linewidth}{!}{
\begin{tabular}{l|c}
\toprule
\textbf{Parameters} & \textbf{\# of Inputs = 8, Bit precision = 4} \\ \midrule
Latency (ns)      & 105.6      \\ 
Throughput (GOPS) & 1.8        \\ 
Operating Frequency (GHz)        & 1.81    \\ \bottomrule
\end{tabular}}
\end{table}

\begin{figure*}[t]
		\centering
		\subfigure[Number of cycles]{\includegraphics[scale=0.27]{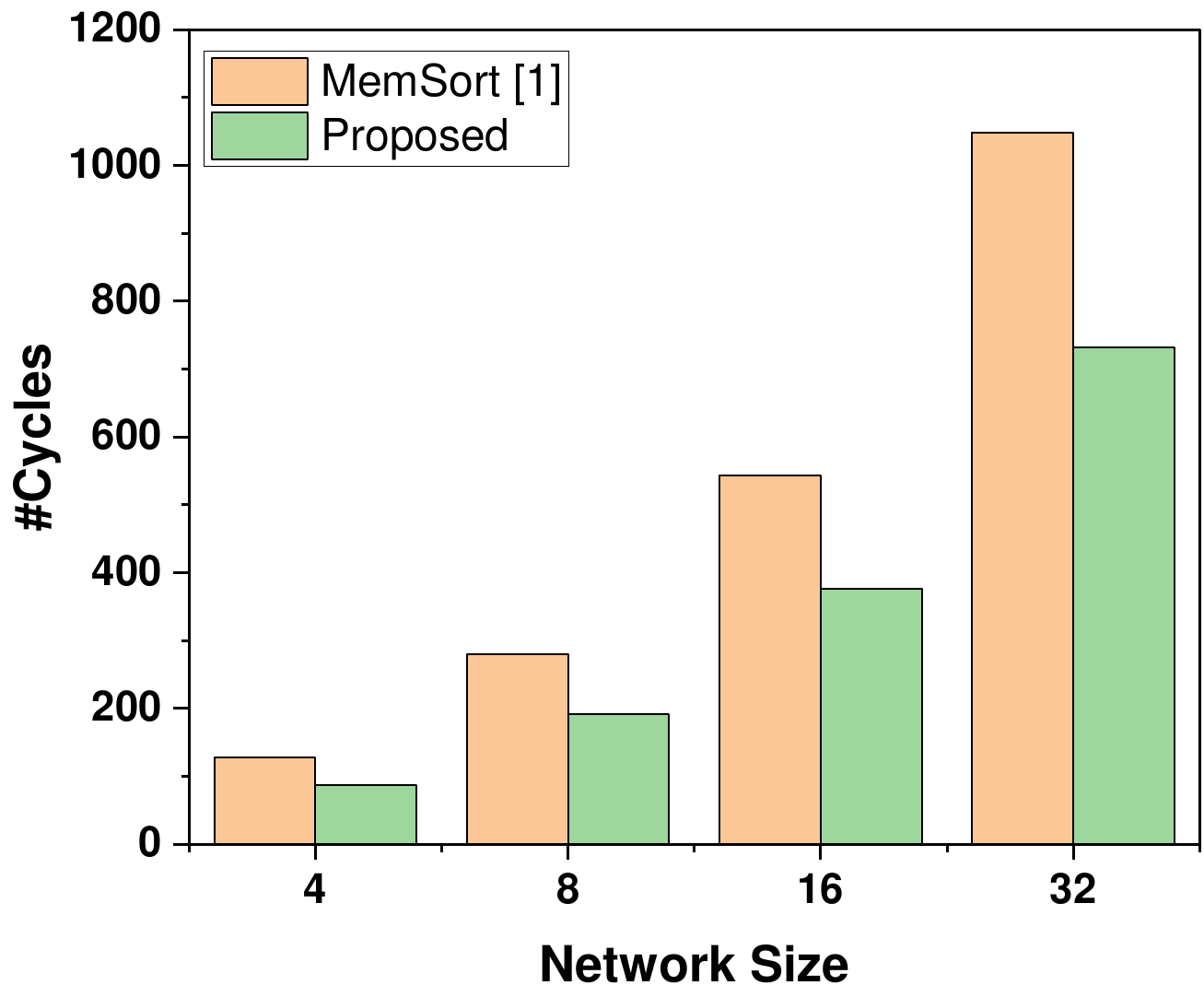}}
		\subfigure[Latency]{\includegraphics[scale=0.27]{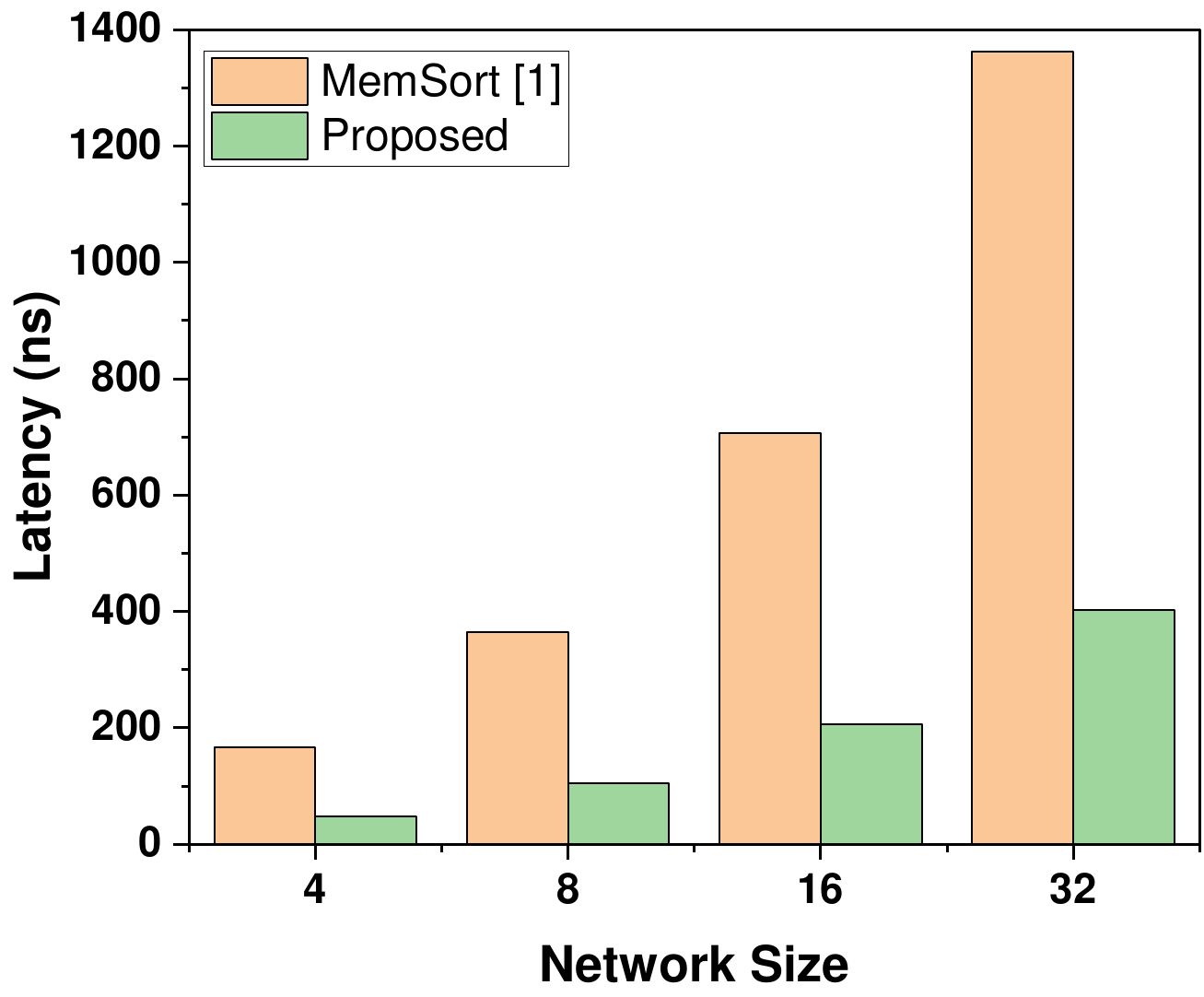}}
		\subfigure[Memory requirement]{\includegraphics[scale=0.27]{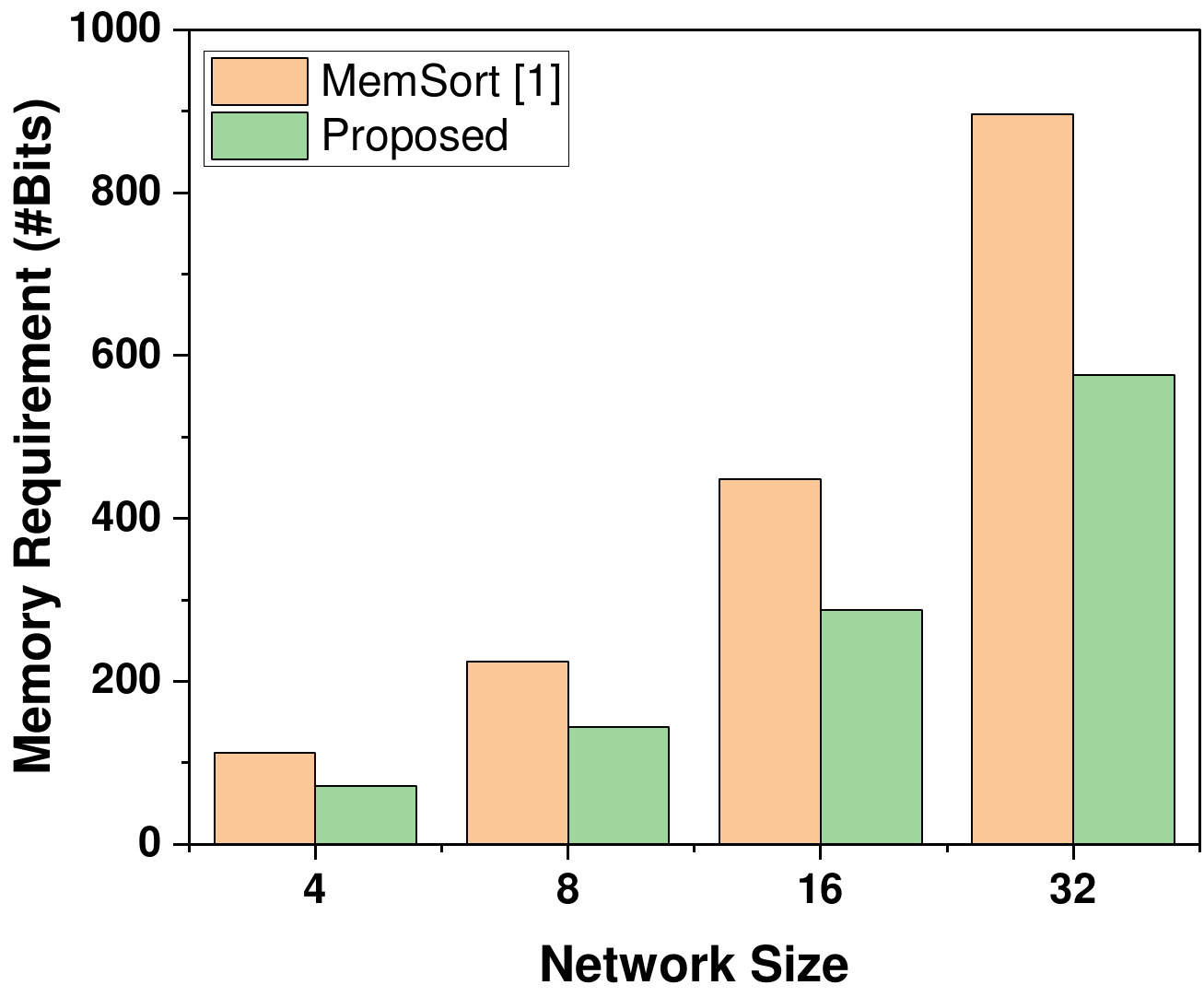}}
		\caption{Performance comparison data sorting of 4-bit for our proposed design and MemSort \cite{Memristor}}
		\label{comparison}
\end{figure*}

We thoroughly compared our proposed architecture's latency against both the bitonic sorting unit and the off-memory computing approach presented in \cite{Memristor}. Also, for software-based sorting, we masked the 8-bit number to 4-bit to sort the data by implementing the bubble sorting algorithm. For these comparisons, we assumed that the initial unsorted data was already stored in SRAM, eliminating the need to include the data loading time in the latency calculations. Our findings demonstrate a significant performance advantage, with our architecture achieving a 5$\times$ reduction in latency compared to memristor-based sorting \cite{Memristor}. 
The comparison of the number of cycles required to sort 4-bit data is depicted in Figure \ref{comparison}(a), highlighting a notable reduction of approximately 1.45$\times$ with the proposed approach. This demonstrates the enhanced efficiency of the design in terms of sorting speed. Similarly, Figure \ref{comparison}(b) illustrates a substantial reduction in latency, with the proposed architecture achieving a 3.4$\times$ improvement. This reduction in latency directly contributes to faster overall processing times.

Figure \ref{comparison}(c) shows the memory utilization for sorting, where the number of memory bits required was minimized by optimizing array utilization. This was achieved by reusing the architecture's rows for the compare-and-swap (CAS) block, leading to a more compact and efficient design.

\section{Conclusion} \label{conclusion}
In this paper, we present a Batcher-based in-memory sorting architecture that leverages the unique characteristics of 6T SRAM to achieve efficient sorting. Integrating CAS blocks directly within the memory array eliminates the von Neumann bottleneck, enabling faster and more energy-efficient sorting operations. The use of SRAM for in-memory computation offers significant advantages over traditional methods, making it a promising approach for data-intensive applications where speed and efficiency are paramount.

As data volumes continue to grow, the need for efficient sorting methods will only increase, driving the development of innovative hardware-based solutions.
In today’s period of AI, data processing is at the center of various applications.
Organizing this data to improve the speed of processing is important, this is one of the examples where sorting can play a crucial part. Our proposed architecture represents a significant step forward in this direction, offering a scalable and efficient solution for sorting in modern computing environments.


\bibliographystyle{IEEEtran}
\bibliography{Reference}

@inproceedings{database1,
title={GPUTeraSort: high performance graphics co-processor sorting for large database management},
author={Govindaraju, Naga and Gray, Jim and Kumar, Ritesh and Manocha, Dinesh},
booktitle={Proceedings of the 2006 ACM SIGMOD internationaInternational Conferencel conference on Management of data},
pages={325--336},
year={2006}
}

@article{database2,
author = {Graefe, Goetz},
title = {Implementing sorting in database systems},
year = {2006},
issue_date = {2006},
publisher = {Association for Computing Machinery},
address = {New York, NY, USA},
volume = {38},
number = {3},
issn = {0360-0300},
doi = {10.1145/1132960.1132964},
journal = {ACM Comput. Surv.},
month = sep,
pages = {10–es},
numpages = {37},
}

@article{scheduling,
title={Implementing scheduling algorithms in high-speed networks},
  author={Stephens, Donpaul C and Bennett, Jon CR and Zhang, Hui},
  journal={IEEE Journal on Selected Areas in Communications},
  volume={17},
  number={6},
  pages={1145--1158},
  year={1999},
  publisher={IEEE}
}

@article{scientific,
title={A novel sorting algorithm and its application to a gamma-ray telescope asynchronous data acquisition system},
  author={Colavita, Alberto and Mumolo, Enzo and Capello, Gabriele},
  journal={Nuclear Instruments and Methods in Physics Research Section A: Accelerators, Spectrometers, Detectors and Associated Equipment},
  volume={394},
  number={3},
  pages={374--380},
  year={1997},
  publisher={Elsevier}
}

@article{FPGA1,
 title={Computer generation of high throughput and memory efficient sorting designs on FPGA},
  author={Chen, Ren and Prasanna, Viktor K},
  journal={IEEE Transactions on Parallel and Distributed Systems},
  volume={28},
  number={11},
  pages={3100--3113},
  year={2017},
  publisher={IEEE}
}

@inproceedings{FPGA2,
  title={FPGASort: A high performance sorting architecture exploiting run-time reconfiguration on FPGAs for large problem sorting},
  author={Koch, Dirk and Torresen, Jim},
  booktitle={Proceedings of the 19th ACM/SIGDA international symposium on Field programmable gate arrays},
  pages={45--54},
  year={2011}
}

@inproceedings{batcher,
  title={Sorting networks and their applications},
  author={Batcher, Kenneth E},
  booktitle={Proceedings of the April 30--May 2, 1968, spring joint computer conference},
  pages={307--314},
  year={1968}
}

@article{batcherlatency,
  title={Bitonic sort on a chained-cubic tree interconnection network},
  author={Baddar, Sherenaz W Al-Haj and Mahafzah, Basel A},
  journal={Journal of Parallel and Distributed Computing},
  volume={74},
  number={1},
  pages={1744--1761},
  year={2014},
  publisher={Elsevier}
}

@article{batcherlatency2,
  title={Modular design of high-throughput, low-latency sorting units},
  author={Farmahini-Farahani, Amin and Duwe III, Henry J and Schulte, Michael J and Compton, Katherine},
  journal={IEEE Transactions on Computers},
  volume={62},
  number={7},
  pages={1389--1402},
  year={2012},
  publisher={IEEE}
}

@article{datadepend,
  title={Low-cost sorting network circuits using unary processing},
  author={Najafi, M Hassan and Lilja, David J and Riedel, Marc D and Bazargan, Kia},
  journal={IEEE Transactions on Very Large Scale Integration (VLSI) Systems},
  volume={26},
  number={8},
  pages={1471--1480},
  year={2018},
  publisher={IEEE}
}

@article{6TSRAM,
    title={IMAC: In-memory multi-bit multiplication and accumulation in 6T SRAM array},
  author={Ali, Mustafa and Jaiswal, Akhilesh and Kodge, Sangamesh and Agrawal, Amogh and Chakraborty, Indranil and Roy, Kaushik},
  journal={IEEE Transactions on Circuits and Systems I: Regular Papers},
  volume={67},
  number={8},
  pages={2521--2531},
  year={2020},
  publisher={IEEE}
}

@article{MainSRAM,
    title={In-Memory Computing with 6T SRAM for Multi-operator Logic Design},
  author={Dhakad, Narendra Singh and Chittora, Eshika and Raut, Gopal and Sharma, Vishal and Vishvakarma, Santosh Kumar},
  journal={Circuits, Systems, and Signal Processing},
  volume={43},
  number={1},
  pages={646--660},
  year={2024},
  publisher={Springer}
}

@inproceedings{CAS,
  title={High performance sorting on the Cell processor [C]},
  author={Gedik, B and Bordawekar, RR and Yu, P S Cellsort},
  booktitle={Proceedings of the 33rd International Conference on Very Large Date Bases, Vienna, Austria},
  pages={52--60},
  year={2009}
}

@inproceedings{Memorypartition,
  title={Felix: Fast and energy-efficient logic in memory},
  author={Gupta, Saransh and Imani, Mohsen and Rosing, Tajana},
  booktitle={2018 IEEE/ACM International Conference on Computer-Aided Design (ICCAD)},
  pages={1--7},
  year={2018},
  organization={IEEE}
}

@article{Softwarealgo,
  title={Review on sorting algorithms a comparative study},
  author={Al-Kharabsheh, Khalid Suleiman and AlTurani, Ibrahim Mahmoud and AlTurani, Abdallah Mahmoud Ibrahim and Zanoon, Nabeel Imhammed},
  journal={International Journal of Computer Science and Security (IJCSS)},
  volume={7},
  number={3},
  pages={120--126},
  year={2013}
}

@article{Memristor,
  title={Sorting in memristive memory},
  author={Alam, Mohsen Riahi and Najafi, M Hassan and TaheriNejad, Nima},
  journal={ACM Journal on Emerging Technologies in Computing Systems (JETC)},
  volume={18},
  number={4},
  pages={1--21},
  year={2022},
  publisher={ACM New York, NY}
}

@ARTICLE{vishal,
  author={Sharma, Vishal and Kim, Ju-Eon and Kim, Hyunjoon and Lu, Lu and Kim, Tony Tae-Hyoung},
  journal={IEEE Journal on Emerging and Selected Topics in Circuits and Systems}, 
  title={A Reconfigurable 16Kb AND8T SRAM Macro With Improved Linearity for Multibit Compute-In Memory of Artificial Intelligence Edge Devices}, 
  year={2022},
  volume={12},
  number={2},
  pages={522-535},
  doi={10.1109/JETCAS.2022.3168571}}

\newpage
\pagestyle{empty}

\end{document}